%% file: PIMRC_CellFree_mmWave_Alonzo2.tex
\documentclass[conference]{IEEEtran}

\IEEEoverridecommandlockouts
\usepackage[letterpaper, left=1in, right=1in, bottom=1in, top=0.75in]{geometry}

\usepackage{ucs}
\usepackage[utf8x]{inputenc}
\usepackage[cmex10]{amsmath}
\usepackage{cite, amsfonts, amssymb, amsthm, bm, bbm, graphicx, relsize, multirow, booktabs, tikz,subfigure,soul}
\usepackage[american]{babel}
\usepackage[T1]{fontenc}
\usepackage{algorithmic, algorithm,color}
\usepackage[multiple]{footmisc}
\usepackage{glossaries}
\theoremstyle{definition}

\theoremstyle{remark}

\input{acronyms}

\usepackage{cite}

\ifCLASSINFOpdf
\else
\fi
\usepackage[cmex10]{amsmath}

\usepackage{algorithmic}

\IEEEoverridecommandlockouts \IEEEpubid{\makebox[\columnwidth]{ 978-1-5386-3531-5/17/\$31.00~\copyright~2017 IEEE \hfill} \hspace{\columnsep}\makebox[\columnwidth]{ }}

\begin{document}
\title{Cell-Free and User-Centric Massive MIMO at Millimeter Wave Frequencies}

\author{\IEEEauthorblockN{Mario Alonzo, and Stefano Buzzi}
\IEEEauthorblockA{Department of Electrical and Information Engineering\\
University of Cassino and Lazio Meridionale, Cassino, Italy}
}


%


\maketitle

\begin{abstract}
In a cell-free (CF) massive MIMO architecture a very large number of distributed access points (APs) simultaneously and jointly serves a much smaller number of mobile stations (MSs); a variant of the cell-free technique is the user-centric (UC) approach, wherein each AP just decodes a reduced set of MSs, practically the ones that are received best. This paper introduces and analyzes the CF and UC architectures at millimeter wave (mmWave) frequencies. First of all, a multiuser clustered channel model is introduced in order to account for the correlation among the channels of nearby users; then, an uplink multiuser channel estimation scheme is described along with low-complexity hybrid analog/digital beamforming architectures. Interestingly, in the proposed scheme no channel estimation is needed at the MSs, and the beamforming schemes used at the MSs are channel-independent and have a very simple structure. Numerical results show that the considered architectures provide good performance, especially in lightly loaded systems, with the UC approach  outperforming the CF one. 
\end{abstract}


%
\IEEEpeerreviewmaketitle

\section{Introduction}
Future fifth-generation (5G) wireless systems will heavily rely on the use of large-scale antenna arrays, a.k.a. massive MIMO, and of carrier frequencies above 10GHz, the so called mmWave frequencies \cite{WhatWill5gbe}. 
For conventional sub-6 GHz frequencies, a new communications architecture, named ``Cell Free'' (CF) massive MIMO, has been recently introduced in \cite{CFNgo1,CFNgo2}, in order to alleviate the cell-edge problem and thus increase the system performance of unlucky users that happen to be located very far from their serving access point (AP). 
In the CF architecture, instead of having few base stations with massive antenna arrays, a very large number of simple APs randomly and densely deployed serve a much smaller number of mobile stations (MSs). In the CF architecture described in \cite{CFNgo1,CFNgo2}, single-antenna APs and MSs are considered, all the APs serve all the MSs, all the APs are connected through a backhaul link to a central processing units (CPU), but every AP performs locally channel estimation, and channel estimates are not sent to the CPU, but are locally exploited. 
In \cite{WSA2017_buzzi_dandrea_CF} the CF architecture is generalized to the case in which both the APs and the MSs are equipped with multiple antennas and, mostly important a user-centric (UC) variant of the CF approach is introduced, wherein each APs, instead of serving all the MSs in the considered area, just serves the ones that he receives best. The results in \cite{WSA2017_buzzi_dandrea_CF} show that the UC approach provides savings on the required backhaul capacity and, also, provides better data-rates to the vast majority of the users. 

This paper, to the best of our knowledge, is the first to consider the CF and UC architectures for mmWave frequencies. The contribution of the paper can be summarized as follows. First of all, we introduce a multiuser mmWave channel model that permits taking into account channel correlation for close users. Building upon the well-known clustered channel model \cite{buzzidandreachannel_model} widely used at mmWave frequencies, we extend this model to take into account the fact that if several APs and MSs are in the same area, their channels must be build using the same set of scatterers; adopting this model, users that are very close will receive beams with very close direction of arrival and so channel correlation for nearby users is intrinsically taken into account. Then, we study the UC and CF approaches at mmWave frequencies; we assume that both the APs and MSs are equipped with multiple antennas, use hybrid analog-digital partial zero-forcing beamforming at the APs, while a very simple 0-1 beamforming architecture, independent of the channel estimate, is used at the MSs. Our results are encouraging since they show that, in the region of interest for the transmitted powers, the considered system is able to provide good performance, with the UC approach outperforming the CF one. In particular,  in a lightly loaded system, taking into account channel estimation errors and using low-complexity beamforming structures, the downlink (uplink) achievable rate-per-user is about 800 (200) Mbit/s, using a bandwidth of 200 MHz. For heavily loaded systems, instead, some performance degradation is observed, and thus more sophisticated beamforming structures and transceiver algorithms are needed to restore the system performance.

This paper is organized as follows. Next Section is devoted to the discussion of the used system and channel model; Section III deals with the description of the communication protocol and of the data detection and channel estimation algorithms, while Section IV contains numerical results. Finally, concluding remarks are given in Section V.

\section{System Model}
We consider a square area of size $250\times250$ sqm, where $K$ MSs and  $M$ APs are randomly located.
The APs are connected by means of a backhaul network to a central processing unit (CPU) wherein data-decoding is performed. Communications take place on the same frequency band; downlink and uplink are separated through time-division-duplex (TDD)\footnote{In TDD the uplink channel is the reciprocal of downlink channel.}. The communication protocol is made of three different phases: uplink training, downlink data
transmission and uplink data transmission. During the uplink training phase, the MSs send pilot sequences to the APs and each AP estimates the channels; during the second phase the APs use the channel estimates to perform pre-coding and transmit the data symbols; finally, in the third phase the MSs send uplink data symbols to the APs. 
While in the CF approach all the APs simultaneously serve all the MSs (a fully-cooperative scenario), in the UC approach each  AP serves a pre-determined number of MSs, say $N$, and in particular the ones that he receives best.

\subsection{Channel model}
We assume that each AP (MS) is equipped with a uniform linear array (ULA) with $N_{AP}$ ($N_{MS}$) elements. 
The ($N_{AP}\times N_{MS}$)-dimensional matrix $H_{k,m}$ denotes the channel matrix between the $k$-th user and the $m$-th AP. According to the widely used clustered channel model for mmWave frequencies (see \cite{buzzidandreachannel_model} and references therein), $\textbf{H}_{k,m}$ can be expressed as 
\begin{equation}
\begin{array}{lll}
\textbf{H}_{k,m}\!\!= &\!\gamma\!\sum\limits_{i=1}^{N_{cl}}\!\sum\limits_{l=1}^{N_{ray}}\!\! \alpha_{i,l}\sqrt{L(r_{i,l})}\textbf{a}_{AP}(\theta_{i,l,k,m}^{AP})\textbf{a}_{MS}^{H}(\theta_{i,l,k,m}^{MS}) \\ &+ \textbf{H}_{LOS},
\end{array}
\label{eq:HKM}
\end{equation}
where $N_{cl}$ is the number of clusters, $N_{ray}$ is the number of the rays that we consider for each cluster, $\gamma$ is a normalization factor defined as $\sqrt{\dfrac{N_{AP}N_{MS}}{N_{cl}N_{ray}}}$, 
$\textbf{H}_{LOS}$ is the line-of-sight (LOS) component, $\alpha_{i,l}$ is the complex path gain distributed as $\mathcal{CN}(0,\sigma^2)$ where $\sigma^2=1$, $L(r_{i,l})$ is the attenuation related to the path $(i,j)$, $\textbf{a}_{AP}$ and $\textbf{a}_{MS}$ are the ULA array responses at the $m$-th AP and at the $k$-th MS, respectively,  and they depend on the angles of arrival and departure, $\theta_{i,l,k,m}^{AP}$ and $\theta_{i,l,k,m}^{MS}$, relative to the 
$(i,l)$-th path of the channel between the $k$-th MS and the $m$-th AP.
The path-loss is defined as \cite{haneda20165g}
\begin{equation}
\label{attenuation_il}
L(r)=-20\log_{10}\biggl(\frac{4\pi}{\lambda}\biggr)-10n\biggl[1+\dfrac{bc}{\lambda f_0}\biggr]\log_{10}(r)-X_{\sigma},
\end{equation}
wherein $n$ is the path loss exponent, $X_\sigma$ is the shadow fading term in logarithmic units with zero mean and $\sigma^2$-variance, and $f_0$  is a fixed frequency (see also table \ref{table:los}). The 
$\textbf{H}_{LOS}$ in (\ref{eq:HKM}) is written as\footnote{For the ease of notation we omit the pedices $k,m$.} 
$
\textbf{H}_{LOS}=I(d)\sqrt{N_{AP}N_{MS}}e^{j\eta}\sqrt{L(d)}\textbf{a}_{AP}(\theta_{LOS}^{AP})\textbf{a}_{MS}^{H}(\theta_{LOS}^{MS}).
$
In the above equation, $\eta\sim\mathcal{U}(0,2\pi)$, $I(d)$ is a 0-1 random variate indicating if a LOS link exists between the transmitter and the receiver, and $d$ is the link length. Denoting by $p$ the probability that $I_{LOS}(d)=1$, we have, for the  UMi (Urban Microcellular) scenarios: 
$
p=\min\biggl(\frac{20}{d},1\biggr) (1-e^{-\frac{d}{39}})+e^{-\frac{d}{39}}. 
$ 
\begin{table}[t] 	
	\center  	 
	\begin{tabular}{|c|c|}
		\hline 
		Scenario & Model Parameters \\ 
		\hline 
		UMi Street Canyon LOS & n=1.98, $\sigma$=3.1dB \\
		\hline 
		UMi Street Canyon NLOS & n=3.19, $\sigma$=8.2dB \\ 
		\hline 
		UMi Open Square LOS & n=2.89, $\sigma$=7.1dB \\ 
		\hline 
		UMi Open Square NLOS & n=1.73, $\sigma$=3.02dB  \\ 
		\hline 
	\end{tabular}  
	\caption{Parameters for Pathloss model}
	\label{table:los}
\end{table}
So far, nothing has been said about the number of scatterers and their positions. While usually for every APMS pairs, a random set of scatterers is considered to contribute to the channel matrix \eqref{eq:HKM}, in this paper, in order to model the possible channel correlation when the devices are closely spaces, we consider the same set of scatterers for the generation of all the channels. In particular, we assume that, in the considered $250 \times 250$ sqm. area, there are a total of 25.000 clusters (corresponding to a cluster density of 0.4 cluster/sqm.), and each cluster contributes with 3 rays. Given these clusters, in order to generate the generic channel $\textbf{H}_{k,m}$ between the $k$-th MS and the $m$-th AP, we consider as active only those clusters falling in an ellipse built arounf the position of  the MS and the AP: this way we exclude far clusters from contributing to the channel. Further details on the multiuser channel generation procedure are omitted due to lack of space and will be reported in an extended version of this paper.


\section{The communication protocol}
In the following, we assume that each MSs employs a very simple 0-1 beamforming structure; in particular, 
 the $(N_{MS}\times P)$-dimensional beamformer user at the $k$-th MS is denoted by $\textbf{L}_k$ and is defined as
 $\textbf{L}_k=\textbf{I}_P\otimes \textbf{1}_{N_{MS}/P}$, with $\otimes$ denoting Kronecker product and $\textbf{1}_{N_{MS}/P}$ an all-1 vector of length $N_{MS}/P$. Otherwise stated, we assume that the MS receive antennas are divided in $P$ disjoint groups of $N_{MS}/P$ elements, and the data collected at the antennas of each group are simply summed together. It is APs' task, based on the uplink channel estimates and exploiting the TDD channel reciprocity, to ensure that the summed samples are, at least approximately, aligned in phase. We describe now the three phases of the communication protocol.

\subsection{Uplink training}
During the uplink training the MSs transmit pilot sequences in order to enable channel estimation at the APs. Let 
$\tau_c$ be the length  of the channel coherence time and $\tau_p$ be the length of uplink training phase, both in discrete time samples.  Of course we must have $\tau_p < \tau_c$. We define by $\boldsymbol{\Phi}_k \in \mathcal{C}^{P\times \tau_p}$ the matrix containing on its rows the pilot sequences sent by the $k$-th MS. We assume that 
$\boldsymbol{\Phi}_k\boldsymbol{\Phi}_k^H =\textbf{I}_P$,  i.e. the rows of $\boldsymbol{\Phi}_k$ are orthogonal, but no orthogonality is required for the pilot sequences assigned to other MSs\footnote{Of course, when $KP\leq \tau_p$ it would be possible to assign to all the MSs mutually orthogonal pilot sequences. In this paper, however, we assume that the pilot sequences are binary random sequences, and we just require that each matrix $\boldsymbol{\Phi}_k$ has orthogonal rows.}. 
  The received signal at the $m$-th AP in the $\tau_p$ signaling intervals devoted to uplink training can be cast in the following  $N_{AP}\times \tau_p$-dimensional matrix $\textbf{Y}_m=\sum\limits_{k=1}^{K}\sqrt{p_k}\textbf{H}_{k,m}\textbf{L}_{k}\boldsymbol{\Phi}_k+\textbf{W}_m$,
where $\textbf{W}_m$ is the matrix of thermal noise samples, whose entries are assumed to be i.i.d. $\mathcal{CN}(0,\sigma_w^2)$ RVs. Letting now $\textbf{S}_{k,m}=\textbf{H}_{k,m}\textbf{L}_{k}$, at the $m$-th AP , an estimate for the quantities $\{\textbf{S}_{k,m}\}_{k=1}^{K}$ can be obtained as follows: 
\begin{equation}
\begin{split}
	\widehat{\textbf{S}}_{k,m}=&\frac{1}{\sqrt{p_k}}\textbf{Y}_{m}\boldsymbol{\Phi}_k^H=\textbf{H}_{k,m}
	\textbf{L}_{k} +\\&\sum\limits_{l=1,l\neq k}^{K}\sqrt{\frac{p_l}{p_k}}\textbf{H}_{l,m}\textbf{L}_{l}\boldsymbol{\Phi}_l\boldsymbol{\Phi}_k^H+\frac{1}{\sqrt{p_k}}\textbf{W}_m\boldsymbol{\Phi}_k^H \; .
	\end{split}
\end{equation}
The estimation must be performed in all APs, i.e. for all $m=1,\dots,M$ and for all $k=1,\dots,K$. Of course, more sophisticated channel estimation schemes can be applied but here we are targeting an extremely simple system processing.

\subsection{Downlink data transmission}
After the first phase, the generic $m$-th AP has an estimate of the quantities $\textbf{S}_{k,m}$, for all $k=1, \ldots, K$. 
In order to transmit data on the downlink a zero-forcing precoder is considered. In particular, denoting by $\textbf{Q}_{k,m}$ the beamformer at the $m$-th AP for transmitting at the $k$-th MS, we require that 
$\widehat{S}_{k,m}^H \textbf{Q}_{k,m} = \textbf{I}_P$ and that $\widehat{S}_{j,m}^H \textbf{Q}_{k,m}$ is zero, for all $j \neq k$. Of course these conditions can be verified only if the number of interfering directions $KP-1$ is smaller than the number of antennas at the APs. When this condition is not fulfilled perfect interference cancellation cannot be achieved. 

The previosuly described beamforming matrix is a fully-digital (FD) one, which presumes the use of a number of RF chains equal to the number of transmit antennas. It is well-known that at mmWave frequencies hardware complexity constraints usually prevent the use of FD architectures, and thus hybrid (HY) beamforming structures have been proposed. 
In this paper we exploit the \emph{Block Coordinate Descent algorithm} \cite{Hybrid_BCDSD} in order to decompose our beamformer in the cascade of a FD one, represented by a $(P \times P)$-dimensional matrix and of a analog one, represented by a $(N_{AP}\times P)$-dimensional matrix whose entries have all constant norm. At the generic $m$-th AP, we will have as many digital beamformers as the MSs to transmit to, and only one analog beamformer, that will be used to transmit jointly to all the users.

\subsubsection{The CF approach} In this case, all the APs can communicate with all the MSs, so the transmitted signal from the \emph{m}-th AP in the \emph{n}-th sample interval is
$\textbf{s}_{m}^{CF}(n)=\sum\limits_{k=1}^{K}\sqrt{\eta_{m,k}^{DL,CF}}\textbf{Q}_{k,m}^{DL} \textbf{x}_{k}^{DL}(n),
$ where $\textbf{x}_{k}^{DL}(n)$ is the data symbol intended for the $k$-th MS, and  $\eta_{m,k}^{DL,CF}$ is a scalar coefficient taking into account the transmit power and it is defined as
$
\eta_{m,k}^{DL,CF}=\dfrac{P_T}{K tr(\textbf{Q}_{k,m}^{DL}(\textbf{Q}_{k,m}^{DL})^{H})}. 
$ 
Note that we are assuming here that, similarly to \cite{WSA2017_buzzi_dandrea_CF}, each AP uniformly divides its power among all the users, and the use of power control rule is left for future work.
The $k-th$ MS receives the following ($N_{MS} \times 1$)-dimensional vector: 
\begin{equation}
\label{eq:r_k}
\begin{split}
\textbf{r}_{k}^{CF}(n)= &\sum\limits_{m=1}^{M}\textbf{H}_{k,m}^{H}\textbf{s}_{m}^{CF}(n) + \textbf{z}_{k}(n)=\\ =&\! \sum\limits_{m=1}^{M}\!\!\!\sqrt{\eta_{m,k}^{DL,CF}}\textbf{H}_{k,m}^{H}\textbf{Q}_{k,m}^{DL} \textbf{x}_{k}^{DL}(n)\!+\!\!\!\!\!\sum\limits_{l=1, l\neq{k}}^{K}\!\sum\limits_{m=1}^{M}\\
&\sqrt{\eta_{m,l}^{DL,CF}}\textbf{H}_{k,m}^{H}\textbf{Q}_{l,m}^{DL} \textbf{x}_{l}^{DL}(n) + \textbf{z}_{k}(n),
\end{split}		
\end{equation}
where $\textbf{z}_{k}(n)$ is the additive thermal noise distributed as $\mathcal{CN}(0,\sigma_{z}^2)$.
A  soft estimate of the $k$-th MS data symbol is thus formed as 
$
\widehat{\textbf{x}}_{k}^{DL,CF}(n)=\textbf{L}_{k}^{H}\textbf{r}_{k}^{CF}(n).
$

\subsubsection{The UC approach} In this case  the APs are assumed to serve a pre-determined, fixed number of MSs, say $N$; in particular, we assume that the generic $m$-th AP serves the $N$ MSs whose channels have the largest Frobenious norms.  We denote by $\mathcal{K}(m)$ the set of MSs served by the $m$-th AP. Given the sets $\mathcal{K}(m)$, for all $m = 1,\dots,M$, we can define the set $\mathcal{M}(k)\triangleq\{m:k\in\mathcal{K}(m)\}$ of the APs that communicate with the $k$-th user.
In this case, the transmitted signal from the $m$-th AP is written as
$\textbf{s}_{m}^{UC}(n)=\sum\limits_{k\in\mathcal{K}(m)}\sqrt{\eta_{m,k}^{DL,UC}}\textbf{Q}_{k,m}^{DL}\textbf{x}_{k}^{DL}(n),
$
where $\eta_{m,k}^{DL,UC}$ is now defined as:
\[
\eta_{m,k}^{DL,UC}=
\begin{cases}
\dfrac{{P_{t}}}{\mid\mathcal{K}(m)\mid tr(\textbf{Q}_{k,m}^{DL}(\textbf{Q}_{k,m}^{DL})^{H})}, &k \in \mathcal{K}(m)\\
0, &k \notin \mathcal{K}(m)
\end{cases}
\]
The received signal at the $k$-th MS is expressed now as: 
\begin{equation}
\label{eq:r_k_Uc}
\begin{array}{ll}
\textbf{r}_{k}^{UC}(n)= \sum\limits_{m=1}^{M}\textbf{H}_{k,m}^{H}\textbf{s}_{m}^{UC}(n) + \textbf{z}_{k}(n)= \\
=\sum\limits_{m\in\mathcal{M}(k)}\sqrt{\eta_{m,k}^{DL,UC}}\textbf{H}_{k,m}^{H}\textbf{Q}_{k,m}^{DL} \textbf{x}_{k}^{DL}(n) + \\
+\sum\limits_{l=1, l\neq{k}}^{K}\sum\limits_{m\in\mathcal{M}(l)}\sqrt{\eta_{m,l}^{DL,UC}}\textbf{H}_{k,m}^{H}\textbf{Q}_{l,m}^{DL} \textbf{x}_{l}^{DL}(n) + \textbf{z}_{k}(n),
\end{array}		
\end{equation}
where the $N_{MS}$-dimensional vector $\textbf{z}_{k}(n)$ represents the thermal noise at the $k$-th MS, and it is modeled as i.i.d. $\mathcal{CN}(0,\sigma_{z}^{2})$. Then it is possible to obtain a soft estimate of the data symbol $\textbf{x}_{k}^{DL}(n)$ at $k$-th MS as
$\hat{\textbf{x}}_{k}^{DL,UC}(n)=\textbf{L}_{k}^{H}\textbf{r}_{k}^{UC}(n).$

\subsection{Uplink data transmission}
The third phase of the communication protocol amounts to uplink data transmission. 
We  denote by $\textbf{x}_{k}^{UL}(n)$ the $P$-dimensional data vector to be transmitted by the $k$-th MS in the \emph{n}-th sample time;
the corresponding signal received at the \emph{m}-th AP is expressed as: 
\begin{equation}
\label{y_m_UL}
\textbf{y}_{m}(n)=\sum\limits_{k=1}^{K}\sqrt{\eta_{k}^{UL}}\textbf{H}_{k,m}\textbf{L}_{k}\textbf{x}_{k}^{UL}(n)+\textbf{w}_m(n)
\end{equation}
where 
$\eta_{k}^{UL}=\dfrac{P_{t}^{UL}}{tr(\textbf{L}_{k}^{H}\textbf{L}_{k})}$,
and $P_{t}^{UL}=1$ is the uplink transmitted power.

\subsubsection{CF approach} in this case, each AP forms the statistic 
$\tilde{\textbf{y}}_{m,k}(n)=\textbf{Q}_{k,m}^H\textbf{y}_{m}(n)$, 
$\forall k$.
Then each AP sends to the CPU the vectors $\tilde{\textbf{y}}_{m,k}(n)$ via the backhaul link, and the CPU forms the following soft estimate of the data vectors transmitted by th $k$-th MS:
\begin{equation}
\label{eq:x_k_hat_UL_Perfect CSI}
\hat{\textbf{x}}_{k}^{UL}(n)=\sum\limits_{m=1}^{M}\tilde{\textbf{y}}_{m,k}(n), \;\;\; k=1,\dots, K. 
\end{equation}
\subsubsection{UC approach} in this case the signal transmitted by the \emph{k}-th MS is decoded only by the APs belonging to the set $\mathcal{M}(k)$. Accordingly, the CPU performs the following soft estimate: 
\begin{equation}
\label{x_k_ul_UC}
\hat{\textbf{x}}_{k}^{UL,UC}=\sum\limits_{m\in\mathcal{M}(k)}\tilde{\textbf{y}}_{m,k}(n), \;\;\; k=1,\dots,K,
\end{equation}

\begin{figure}[t]
\centering
\includegraphics[scale=0.21]{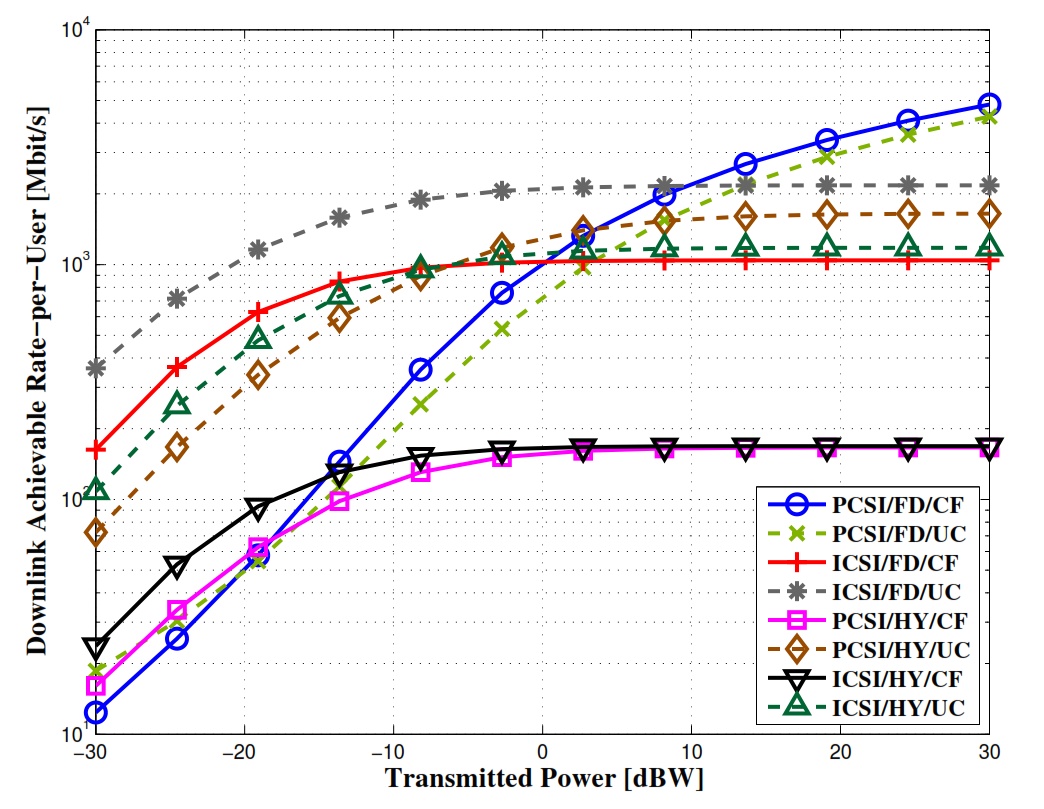}
\caption{Average downlink  achievable rate per user versus transmit power. System parameters: $M=100$, $K=5$, $N_{AP} \times N_{MS}=16 \times 8$, $P=2$, $N=3$.}
\label{Fig:DLK5}
\end{figure}

\begin{figure}[t]
\centering
\includegraphics[scale=0.21]{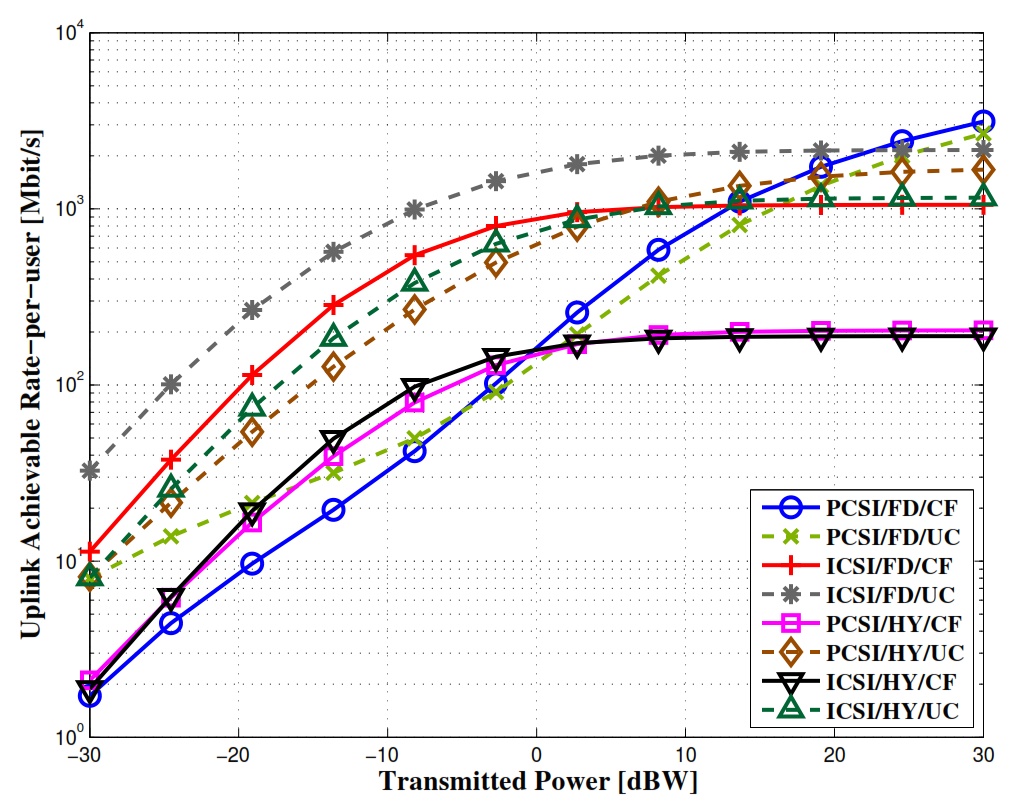}
\caption{Average uplink  achievable rate per user versus transmit power. System parameters: $M=100$, $K=5$, $N_{AP} \times N_{MS}=16 \times 8$, $P=2$, $N=3$.}
\label{Fig:ULK5}
\end{figure}

\begin{figure}[t]
\centering
\includegraphics[scale=0.21]{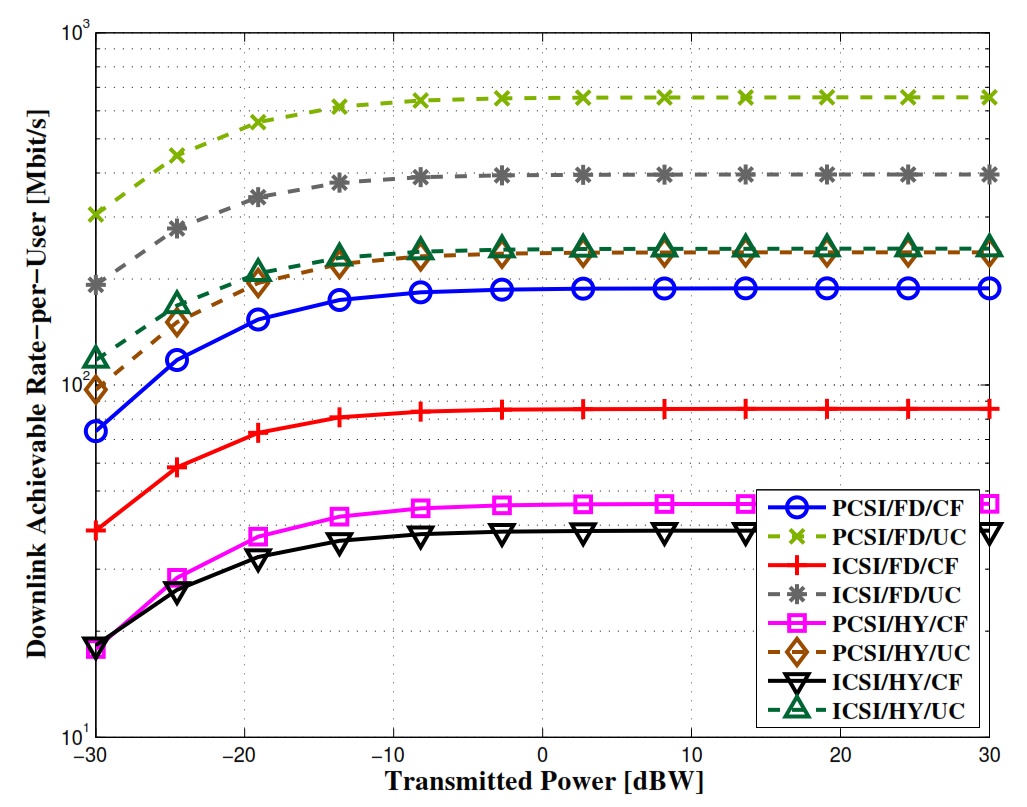}
\caption{Average downlink  achievable rate per user versus transmit power. System parameters: $M=100$, $K=20$, $N_{AP} \times N_{MS}=16 \times 8$, $P=2$, $N=3$.}
\label{Fig:DLK20}
\end{figure}

\begin{figure}[t]
\centering
\includegraphics[scale=0.21]{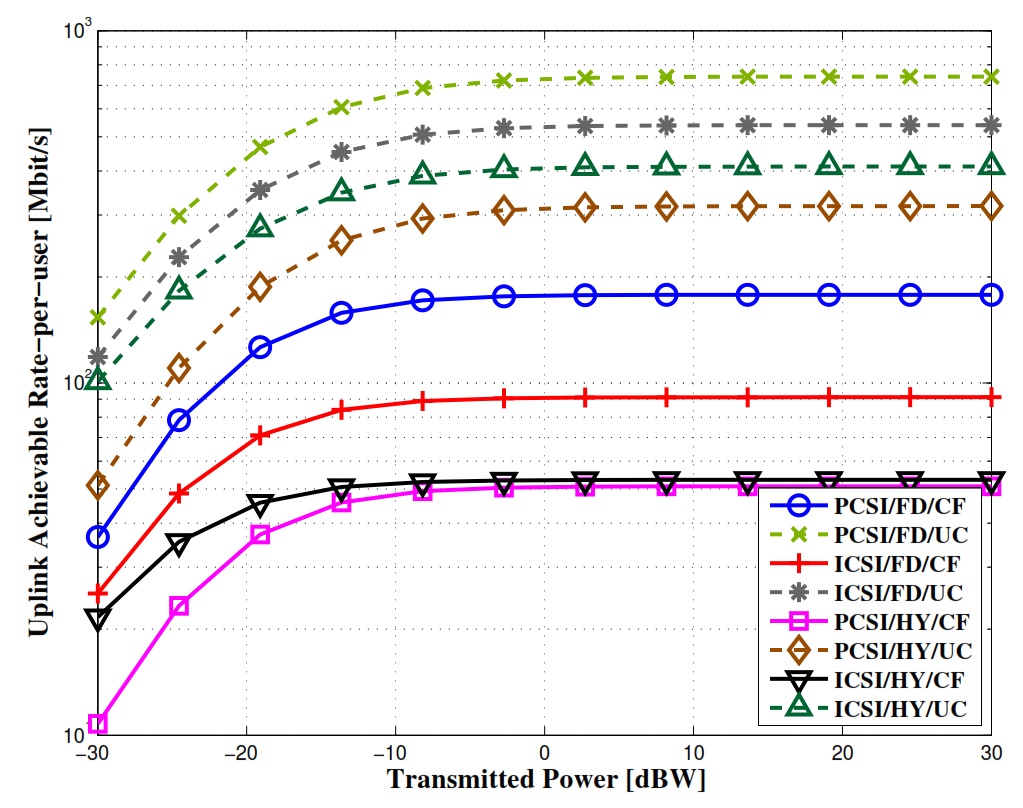}
\caption{Average uplink  achievable rate per user versus transmit power. System parameters: $M=100$, $K=20$, $N_{AP} \times N_{MS}=16 \times 8$, $P=2$, $N=3$.}
\label{Fig:ULK20}
\end{figure}

\section{Numerical Results}
As a measure of system performance we use the average achievable  rate-per-user, measured in Mbit/s.

We consider a carrier frequency $f_0=$ 73 GHz and a bandwidth $B =$ 200MHz. The scenario is the UMi Open Square; 
the additive noise has a power spectral density of -174 dBm/Hz, the receiver noise figure is $F=6$ dB. We assume that there are $M=100$ APs randomly deployed, and each AP has $N_{AP}=16$ antennas. The MSs have $N_{MS}=8$ antennas; a lightly loaded scenario ($K=5$) and a heavy loaded scenario ($K=20$) is considered for performance evaluation. The considered multiplexing order is $P=2$. We show results for the case of imperfect CSI (ICSI) and perfect CSI (PCSI), for the CF case and for the UC case; in the UC case, we have taken $N=1$, for the case $K=5$, and $N=3$, for the case $K=20$. The ICSI results have been obtained considering pilot sequences of length $\tau_p=128$ and an uplink transmit power of 100mW. 
We also show results for the case of FD beamforming and of HY beamforming, with a number of RF chains equal to  the multiplexing order $P$.
Figs. \ref{Fig:DLK5} and \ref{Fig:ULK5} show, considering 60 independent scenario realizations,  the average rate-per-user, on the downlink and on the uplink, versus the transmitted power for the lightly loaded scenario. Figs. \ref{Fig:DLK20} and \ref{Fig:ULK20}  show the same results for the highly loaded scenario. 
Although the considered transmitted power range is $[-30, 30]$ dBW, the region of interest is in the range $[-10, 0]$ dBW for the downlink and $[-20, -10]$ dBW for the uplink. Inspecting the figures, some comments can be done. First of all, there is a saturation effect for all the considered structures, with the exception of the PCSI/FD structures in the lightly loaded case. This can be seen as an indication that the system is interference-limited in the high transmit power region, and this of course comes at no surprise since the PCSI/FD ZF structure is able to get rid of the interference only in the lightly loaded case, while in the highly loaded case the interference subspace fills all the available dimensions. The saturation is also due to the fact that we are using a very simple 0-1 channel-independent beamforming structure at the MSs. Nonetheless, in the previously cited regions of interest for the transmit powers, it is seen that the suboptimal structures relying on imperfect CSI provide satisfactory performance, in some cases outperforming the ones with perfect CSI (again this is due to the fact that we are not considering an optimal approach to beamforming). In particular, for the light-loaded case, at a 1W transmit power, the ICSI/HY/UC structure achieves a downlink rate-per-user of about 1 Gbit/s (corresponding to 25 bit/s/Hz spectral efficiency) and the ICSI/HY/CF structure achieves 160 Mbit/s (corresponding to 4 bit/s/Hz spectral efficiency); the corresponding uplink rate-per-user are 800 Mbit/s and 150 MBit/s, for the UC and the CF approaches, respectively. For the highly loaded scenario ($K=20$, i.e. four times more users), the downlink average rate-per-user is 239 Mbit/s for the ICSI/HY/UC and 37 Mbit/s for the ICSI/HY/CF structure. We see that the spectral efficiency values are approximately constant when passing from the lightly loaded to the highly loaded case. We also see that for the case of ICSI, the UC approach greatly outperforms the CF approach. This is due to the fact that each AP obtains very noisy channel estimates for far MSs, and this its contribution to the transmission and reception phase for far MSs basically adds detrimental noise and interference originated from pilot contamination. Comparing our results with those obtained in \cite{WSA2017_buzzi_dandrea_CF} with reference at sub-6 GHz frequencies, we see that the gap between the UC and the CF approach is at mmWave frequencies even bigger than at sub-6 Ghz frequencies, presumably due to the stronger attenuation at mmWave frequencies, which makes the contribution from far APs practically useless and detrimental. 
A similar discussion can be made for the uplink, and is omitted for the sake of brevity.

\section{Conclusion}
The paper has presented first results on the comparison between the CF and UC approach at mmWave frequencies, taking into account the effect of channel estimation errors, the presence of HY beamforming at the APs and of a very simple 0-1 channel independent beamforming structure at the MSs. Despite these simplifications lead to an interference-limited system,  the proposed techniques are able to achieve remarkable values in terms of data-rates and of spectral efficiency, and the UC approach has been shown to outperform the CF one in practical settings.

\bibliographystyle{IEEEtran}
\bibliography{my_references}

\end{document}

%% file: acronyms.tex
\makeglossaries

\newacronym{see}{SEE}{secrecy energy efficiency}
\newacronym{miso}{MISO}{multiple input single output}
\newacronym{miso-se}{MISO-SE}{multiple input single output single-antenna eavesdropper}
\newacronym{lmmse}{LMMSE}{linear minimum mean square error}
\newacronym{d2d}{D2D}{device-to-device}
\newacronym{p2p}{P2P}{point-to-point}
\newacronym{mac}{MAC}{multiple-access channel}
\newacronym{bc}{BC}{broadcast channel}
\newacronym{ic}{IC}{interference channel}
\newacronym{imac}{IMAC}{interference multiple access channel}
\newacronym{ibc}{IBC}{interference broadcast channel}
\newacronym{mimo}{MIMO}{multiple-input multiple-output}
\newacronym{mimo-me}{MIMO-ME}{multiple input multiple output multiple-antenna eavesdropper}
\newacronym{siso}{SISO}{single-input single-output}
\newacronym{sc}{SC}{single-carrier}
\newacronym{mc}{MC}{multi-carrier}
\newacronym{ofdma}{OFDMA}{orthogonal frequency division multiple access}
\newacronym{af}{AF}{amplify-and-forward}
\newacronym{df}{DF}{decode-and-forward}
\newacronym{cf}{CF}{compress-and-forward}
\newacronym{mwrc}{MWRC}{multi-way relay channel}
\newacronym{pde}{PDE}{partial data exchange}
\newacronym{fde}{FDE}{full data exchange}
\newacronym{iid}{i.i.d.\@}{independent and identically distributed}
\newacronym{awgn}{AWGN}{additive white Gaussian noise}
\newacronym{awg}{AWG}{additive white Gaussian}
\newacronym{sic}{SIC}{successive interference cancellation}
\newacronym{dpc}{DPC}{dirty paper coding}
\newacronym{snr}{SNR}{signal-to-noise ratio}
\newacronym{sinr}{SINR}{signal to interference plus noise ratio}
\newacronym{ber}{BER}{bit error rate}
\newacronym{zf}{ZF}{zero-forcing}
\newacronym{mmse}{MMSE}{minimum mean square error}
\newacronym{sud}{SUD}{single user decoding}
\newacronym{dof}{DoF}{degrees of freedom}
\newacronym{gdof}{GDoF}{generalized degrees of freedom}
\newacronym{nnc}{NNC}{noisy network coding}
\newacronym{dmn}{DMN}{discrete memoryless network}
\newacronym{csi}{CSI}{channel state information}
\newacronym{ee}{EE}{energy efficiency}
\newacronym{ian}{IAN}{treating interference as noise}
\newacronym{snd}{SND}{simultaneous non-unique decoding}
\newacronym{brd}{BRD}{best response dynamics}
\newacronym{br}{BR}{best response}
\newacronym{ne}{NE}{Nash equilibrium}
\newacronym{lhs}{LHS}{left-hand side}
\newacronym{rhs}{RHS}{right-hand side}
\newacronym{gee}{GEE}{global energy efficiency}
\newacronym{wsee}{WSEE}{weighted sum energy efficiency}
\newacronym{wpee}{WPEE}{weighted product energy efficiency}
\newacronym{wmee}{WMEE}{weighted minimum energy efficiency}
\newacronym{kkt}{KKT}{Karush Kuhn Tucker}
\newacronym{pc}{PC}{pseudo-concave}
\newacronym{qc}{QC}{quasi-concave}
\newacronym{ql}{QL}{quasi-linear}
\newacronym{pl}{PL}{pseudo-linear}
\newacronym{spc}{SPC}{strictly pseudo-concave}
\newacronym{sqc}{SQC}{strictly quasi-concave}
\newacronym{lfp}{LFP}{linear fractional problem}
\newacronym{clfp}{CLFP}{concave-linear fractional problem}
\newacronym{ccfp}{CCFP}{concave-convex fractional problem}
\newacronym{mmfp}{MMFP}{max-min fractional problem}
\newacronym{sorp}{SoRP}{sum-of-ratios problem}
\newacronym{porp}{PoRP}{product-of-ratios problem}
\newacronym{qos}{QoS}{quality of service}
\newacronym{evd}{EVD}{eigenvalue decomposition}
\newacronym{svd}{SVD}{singular value decomposition}
\newacronym{skee}{SKEE}{Secret-key energy efficiency}
\newacronym{an}{AN}{artificial noise}